\documentclass[5p,accept]{elsarticle}

\usepackage{lineno,hyperref}
\usepackage{amsmath}
\usepackage{amssymb}
\usepackage{amsfonts}
\usepackage{gensymb}
\usepackage{bm}
\usepackage{xcolor}
\usepackage{booktabs} 
\usepackage{multirow}
\usepackage{tikz}

\usepackage{xcolor}
\usepackage{soul}
\sethlcolor{yellow}

\usepackage{graphicx}
\usepackage{subcaption}
\usepackage{siunitx}
\usepackage{hyperref}

\newcommand{\dt}{\Delta t}

\makeatother

\newcommand\norm[1]{\lVert#1\rVert}
\usepackage{here}

\journal{Journal of \LaTeX\ Templates}

\begin{document}

\begin{frontmatter}

\title{Local particle refinement in terramechanical simulations}


\author[mymainaddress,mysecondaryaddress]{Markus Pogulis}
\ead{markus.pogulis@umu.se}

\author[mymainaddress,mythirdaddress]{Martin Servin}
\ead{martin.servin@umu.se}

\address[mymainaddress]{Ume\aa\ University, SE-90187, Ume\aa, Sweden}
\address[mysecondaryaddress]{BAE Systems H\"agglunds AB, SE-89141, \"Ornsköldsvik, Sweden}
\address[mythirdaddress]{Algoryx Simulation AB, SE-90736, Ume\aa, Sweden}

\begin{abstract}
The discrete element method (DEM) is a powerful tool for simulating granular soils, but its high computational demand often results in extended simulation times. While the effect of particle size has been extensively studied, the potential benefits of spatially scaling particle sizes are less explored. We systematically investigate a local particle refinement method's impact on reducing computational effort while maintaining accuracy. We first conduct triaxial tests to verify that bulk mechanical properties are preserved under local particle refinement. Then, we perform pressure-sinkage and shear-displacement tests, comparing our method to control simulations with homogeneous particle size. We evaluate $36$ different DEM beds with varying aggressiveness in particle refinement. Our results show that this approach, depending on refinement aggressiveness, can significantly reduce particle count by $2.3$ to $25$ times and simulation times by $3.1$ to $43$ times, with normalized errors ranging from $3.4$\% to $11$\% compared to high-resolution reference simulations. The approach maintains a high resolution at the soil surface, where interaction is high, while allowing larger particles below the surface. The results demonstrate that substantial computational savings can be achieved without significantly compromising simulation accuracy. This method can enhance the efficiency of DEM simulations in terramechanics applications.
\end{abstract}


\end{frontmatter}

\newpage

\section{Introduction}
The discrete element method (DEM) is a versatile but computationally intensive method for granular dynamics simulation \cite{cundall1979discrete}. In terramechanics applications, it is an important analysis tool in situations when the soil undergoes large plastic deformations or transitions into a rapid shear flow. However, the strength of DEM also presents its most significant challenge. The necessity of tracking the motion and interactions of a large number of particles results in computationally intensive simulations, particularly when accounting for full vehicle dynamics while traversing a terrain. This underscores the need for efficient computational strategies.

The computational demand is often closely linked to particle size and shape. Employing smaller particles yields higher particle resolution and, consequently, better accuracy \cite{miyai2019influence}. Additionally, the choice of particle shape impacts computational efficiency; spherical shapes allow for more efficient contact detection algorithms compared to polygonal ones, but then lack the effects of true interlocking of particles \cite{coetzee2020calibration}. These advantages come with the trade-off of increased computational demands and extended simulation times.

Historically, various scaling methods have been explored to combat these challenges presented by DEM. One such method is \emph{exact scaling}, also referred to as \emph{geometric similarity}, which scales both particles and geometry by the same amount. This method maintains the same number of particles and has been shown to have no significant impact on computational intensity \cite{coetzee2019particle,feng2009upscaling,feng2014discrete}.
Another approach is the use of \emph{coarse-graining}, also referred to as \emph{pseudo-particles} \cite{ge2017discrete,feng2014discrete}. This method scales up particles of all sizes by the same amount, thereby reducing the total number of particles. A third method is \emph{particle scalping}, or \emph{distribution cut-off}, \cite{ahmadi2023scaling,roessler2016scalability,roessler2018scaling}. This method only scales the smallest particles, below a selected cut-off size, while all larger particles remain their size.

\begin{figure*}[t]
    \centering
    \includegraphics[width=0.9\textwidth]{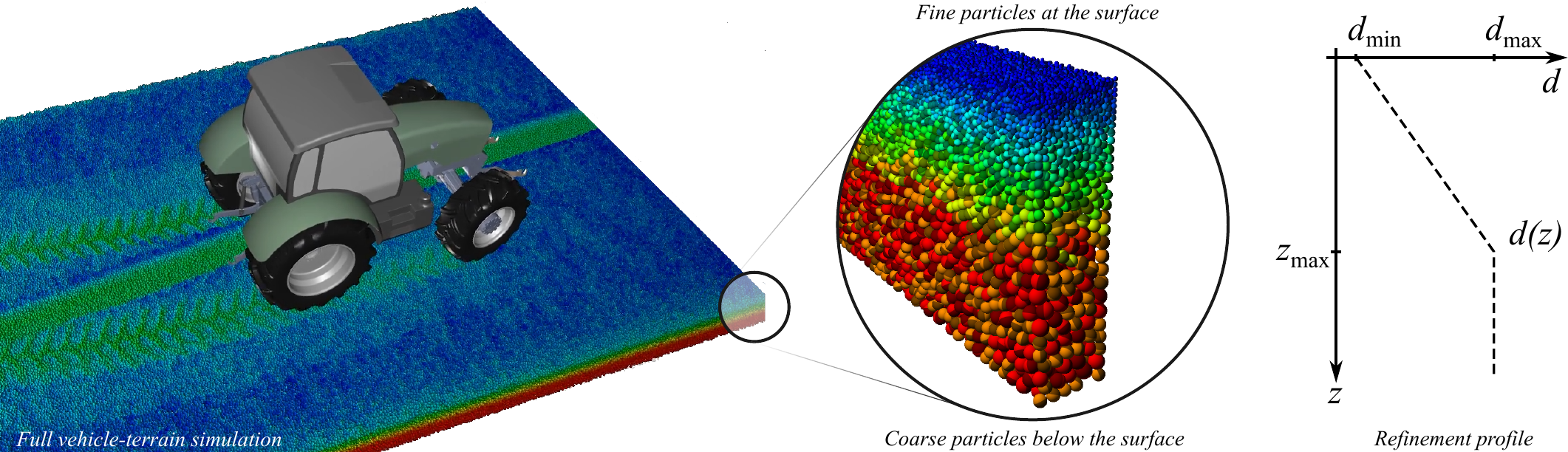}
    \caption{Illustration of the idea of local particle refinement for accelerating vehicle-terrain simulations. Fine particles are used at the terrain surface where they interface with vehicle's tractive devices. Image courtesy of Peter Norlindh at Algoryx Simulation.}
    \label{fig:idea}
\end{figure*}

While each method has its strengths and limitations, they all share some common considerations. Regardless of the method used, model calibration is necessary to ensure that the bulk properties match those of the true particles \cite{coetzee2017calibration}. 
Additionally, when scaling the smallest particles, it's important to recognize that effects at length scales smaller than the smallest particles can no longer be captured. That includes the formation of localized shear bands \cite{rudnicki1975conditions} and their dynamic evolution which have been found to be sensitive to particle size in DEM simulations \cite{miyai2019influence}.

Although the effect of particle size has been extensively investigated in DEM, varying particle size spatially throughout the simulation domain remains largely unexplored. Previously, a few studies have applied such particle refinement in cone penetrometer testing \cite{mcdowell2012particle,sharif2019numerical} and rock mechanics \cite{pan2021attempt}.

Building on these initial efforts, we develop a systematic approach to particle refinement in DEM. The idea is to scale the local particle size (distribution) based on the expected need for spatial resolution. The method is inspired by the use of adaptive mesh refinement in the finite element method (FEM) and other numerical methods for continuum mechanics, where element size is varied based on local error sensitivity. Previous work has demonstrated the potential of particle refinement for cone penetrometer testing but focused primarily on the penetrometer results rather than a systematic investigation of the refinement method itself \cite{mcdowell2012particle,sharif2019numerical}. 
They did find that the number of particles and computational time could be significantly reduced using particle refinement. Also, the effect of small particles from migrating into the voids between
the large particles could be avoided by grading the particles into layers.
Our work extends this approach by examining how different levels of refinement affect accuracy and computational efficiency. 

We hypothesize that the particle refinement method can maintain high accuracy and significantly reduce the total number of particles in the system, thereby reducing the computational load. In a terramechanics application, illustrated in Fig.~\ref{fig:idea}, the method entails using fine particles on the top of the soil bed, where they can properly interact with geometric features such as wheel treads and track grousers, and coarser particles at greater depth. However, it's important to note that this approach may have limitations in applications where there is significant shear flow and mixing between layers of different-sized particles.

To test the method, we create many locally refined soil samples and a few with uniform particle sizes serving as reference cases. We use spherical particles with rolling resistance to model shape and interlocking effects. First, we conduct a series of triaxial tests to verify that the bulk mechanical properties are invariant under local particle refinement. Next, to evaluate the effectiveness and accuracy of particle refinement in a more typical terramechanics application, we simulate both pressure-sinkage and shear-displacement tests. These tests are chosen as they effectively measure the performance of tractive elements depending on their shape and soil strength. All results are compared to similar tests run with uniform particle sizes.

The details of our method and the results of our experiments are presented in the following sections.

\section{Methodology}
\label{sec:modelling}
For discrete element simulation, we use a time-implicit (nonsmooth) version described in Sec.~\ref{sec:DEM}. The technique for local refinement is, however, independent of this choice and therefore presented first.

\subsection{Local particle refinement and scaling agressiveness}
\label{sec:refinement}
Inspired by the use of adaptive mesh refinement in FEM, the idea with local particle refinement in DEM is to use fine-grain particles in regions sensitive to discretization errors and coarser particles further away. Generally, it is difficult to predict where to apply refinement and how much. In terramechanics simulations, it is natural to assume that the particles near the terrain surface need to be fine enough to resolve important features of the tractive element, e.g., thread or grousers on a wheel or track shoe. We denote the smallest particle size by $d_\text{min}$. Then we imagine the particles increasing in size by some function $d(z)$ of the depth $z$. This continues down to some depth $z_\text{max}$ where the particles reach a maximum size $d_\text{max}$. Beyond this depth, the particles remain at their maximum size. For simplicity, we use a linear relation 
$d(z) \propto \gamma z$ and we refer to the constant gradient $\gamma$ as the bed's \emph{scaling aggressiveness}. With this representation, we can describe any particle bed profile by the three numbers $[d_\text{min}, \gamma, d_\text{max}]$

\begin{equation}
    d(z) = \begin{cases}
    d_\text{min} + \gamma z, & \text{for } 0 < z < z_\text{max} \\
    d_\text{max}, & \text{for } z \geq z_\text{max}
\end{cases}
\end{equation}
with $z_\text{max} = (d_\text{max} - d_\text{min})/\gamma$.

A particle bed with a desired scaling aggressiveness can be achieved by creating discrete layers of differently-sized particles. Let each layer, numbered $n = 0,1,2,…$, have thickness $\eta d_n$ and consist of particles with diameter increasing as $d_n = r d_{n-1}$ with constants $r,\eta\geq 1 $. The vertical position of layer $n$ can then be calculated as a geometric sum to $z_n =\sum_{i=0}^n \eta d_i = \eta d_0 \tfrac{r^{n+1}-1}{r-1}$. From this one gets the relation $d_n = \tfrac{d_0}{r} + \tfrac{r-1}{r\eta} z_n$. We identify the discrete scaling  aggressiveness as $\gamma(r,\eta) = \tfrac{r-1}{r\eta}$ and $d_0 = d_\text{min}$. This layer representation is useful for creating beds whether you do it using particle emitters or by placing the particles individually, using e.g. a radial expansion method. It is not limited to layers of mono-sized particles. The method applies to having a particle size distribution within each layer, with $d_n$ being the mean or median diameter.

\begin{figure}
    \centering \includegraphics[width=0.6\columnwidth,keepaspectratio]{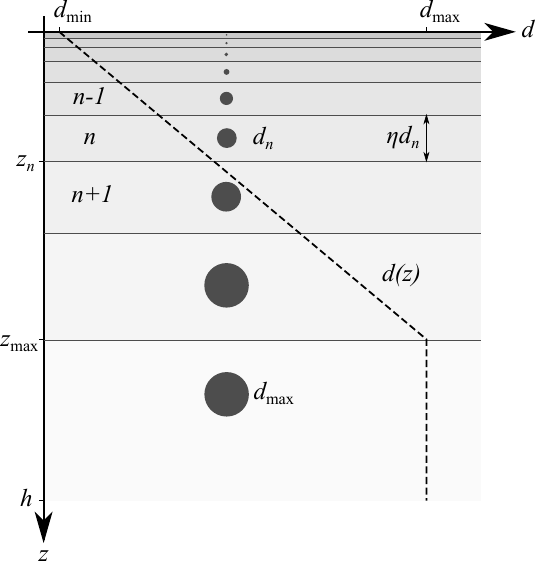}
    \caption{Illustration of a soil bed with vertical particle refinement profile $d(z)$ discretized in layers indexed $n$ with particle size $d_n = r d_{n-1}$ and
    layer thickness $\eta d_n$. Below $z_\text{max}$, the particles have size $d_\text{max}$.}
    \label{fig:scaling_bed}
\end{figure}

\subsection{Nonsmooth DEM and multibody dynamics}
\label{sec:DEM}
We use a time-implicit formulation of DEM and rigid multibody dynamics as described in \cite{servin:2014:esn} and \cite{wiberg2021discrete}, and implemented in AGX Dynamics \cite{AGX2024}.
This is also referred to as a nonsmooth DEM or the contact dynamics method \cite{Radjai:2009:cdn}.
In this framework, there is no fundamental distinction between soil particles and rigid multibodies part of jointed and possibly actuated mechanism. The governing equations are
\begin{align}
  \label{eq:momentum}
  \bm{M} \dot{\bm{v}}  = \bm{f} + \bm{G}_\mathrm{j}^T \bm{\lambda}_\mathrm{j} + \bm{G}_\mathrm{c}^T \bm{\lambda}_\mathrm{c},\\
  \label{eq:kinematic_constraint}
  \varepsilon_{\mathrm{j}} \bm{\lambda}_{\mathrm{j}} + 
        \eta_{\mathrm{j}} \bm{g}_{\mathrm{j}} + 
        \tau_{\mathrm{j}} \bm{G}_{\mathrm{j}} \bm{v} = \bm{u}_{\mathrm{j}}, \\   \label{eq:limits}
    \bm{\lambda}_\mathrm{min} \leq \bm{\lambda}_\mathrm{j} \leq \bm{\lambda}_\mathrm{max},\\  \label{eq:contact_constraint}\mathrm{contact\_law(}\bm{g}_{\mathrm{c}},\bm{v}_{\mathrm{c}},\bm{\lambda}_{\mathrm{c}} \mathrm{)},
\end{align}
where our system of $N_\mathrm{b}$ bodies has velocity $\bm{v}\in\mathbb{R}^{6N_\mathrm{b}}$, mass matrix $\bm{M} \in \mathbb{R}^{6N_\mathrm{b}\times6N_\mathrm{b}}$, and external force $\bm{f}\in\mathbb{R}^{6N_\mathrm{b}}$.
The constraint forces in the Newton-Euler equation of motion (\ref{eq:momentum}), with Lagrange multiplier $\bm{\lambda}$ and Jacobian $\bm{G}$ are divided
joints and motors, labeled with $\mathrm{j}$, and contacts, labeled with $\mathrm{c}$.
Eq.~\eqref{eq:kinematic_constraint} is a generic constraint equation.  An ideal joint can be represented with $\varepsilon_{\mathrm{j}} = \tau_{\mathrm{j}} = \bm{u}_{\mathrm{j}} = 0$, in which case Eq.~\eqref{eq:kinematic_constraint} express a holonomic constraint, $\bm{g}_{\mathrm{j}}(\bm{x}) = 0$.  
A non-ideal joint is modeled using finite compliance $\varepsilon_{\mathrm{j}}$ and viscous damping rate $ \tau_{\mathrm{j}}$.
A linear or angular motor may be represented by a velocity constraint $ \bm{G}_{\mathrm{j}} \bm{v} = \bm{u}_{\mathrm{j}}(t)$ with target speed $\bm{u}_{\mathrm{j}}(t)$, which follows by  $\varepsilon_{\mathrm{j}} = \eta_{\mathrm{j}} = 0$ and $ \tau_{\mathrm{j}} = 1$. Range limits on the motor constraint forces may be imposed by Eq.~\eqref{eq:limits}.
With $N_\mathrm{j}$ constrained and actuated degrees of freedom we have $\bm{\lambda}_{\mathrm{j}}\in\mathbb{R}^{N_\mathrm{j}}$ and $\bm{G}_\mathrm{j}\in\mathbb{R}^{N_\mathrm{j}\times6N_\mathrm{b}}$.

Contact laws are imposed as inequality and complementarity conditions on the contact multiplier $\bm{\lambda}_{\mathrm{c}}\in\mathbb{R}^{3N_\mathrm{c}}$
and relative contact velocity, $\bm{G}_\mathrm{c}\bm{v}$.
Each contact multiplier 
is split $\bm{\lambda}_{\mathrm{c}} = [\lambda_{\mathrm{n}};\bm{\lambda}_{\mathrm{t}};\bm{\lambda}_{\mathrm{r}}]$
in the normal, tangential, and rotational components that must obey the Coulomb condition, 
$|\bm{\lambda}_{\mathrm{t}}| \leq \mu_{\mathrm{t}} \lambda_{\mathrm{n}}$, and analogous rolling friction condition $|\bm{\lambda}_{\mathrm{r}}| \leq (\mu_{\mathrm{r}} d/2) \lambda_{\mathrm{n}}$ with particle diameter $d$. 
The contact laws are imposed as complementarity conditions 
\begin{align}\label{eq:g_n}
    0\leq \varepsilon^{-1}_\mathrm{n}g_\mathrm{n} + \gamma_\mathrm{n} \bm{G}_\mathrm{n}\bm{v} \perp \lambda_\mathrm{n}\geq 0,\\\label{eq:G_t}
    \norm{\bm{G}_\mathrm{t} \bm{v}}(\mu_{\mathrm{t}} \lambda_{\mathrm{n}} - \norm{\bm{\lambda}_{\mathrm{t}}}) = 0,\\ \label{eq:G_r}
    \norm{\bm{G}_\mathrm{r} \bm{v}}(\mu_{\mathrm{r}} r\lambda_{\mathrm{n}} - \norm{\bm{\lambda}_{\mathrm{r}}}) = 0,
\end{align}
for non-penetration, no-slip, no-rolling, respectively. 
Here, $g_\mathrm{n}$ is a contact gap function, with normal Jacobian $\bm{G}_\mathrm{n}=\frac{\partial g_\mathrm{n}}{\partial \bm{x}}$, contact compliance $\varepsilon_\mathrm{n}$ and damping $\gamma_\mathrm{n}$, and the tangential Jacobian $\bm{G}_\mathrm{t}$ is such that the contacting bodies relative velocity in the contact tangent space is given by $\bm{G}_\mathrm{t} \bm{v}$. 
At slip and roll, the direction of the tangential and rolling friction is determined by the principle of maximumum dissipation,
$\norm{\bm{G}_\mathrm{t} \bm{v}}\norm{\bm{\lambda}_{\mathrm{t}}} = - (\bm{G}_\mathrm{t} \bm{v})^T\bm{\lambda}_{\mathrm{t}}$
and $\norm{\bm{G}_\mathrm{r} \bm{v}}\norm{\bm{\lambda}_{\mathrm{r}}} = - (\bm{G}_\mathrm{r} \bm{v})^T\bm{\lambda}_{\mathrm{r}}$.
The set of $N_\mathrm{c}$ active contacts, having a finite contact gap $\delta_\mathrm{c}$, are re-computed at every simulation timestep using a collision detection algorithm. 
Each overlap is translated to a normal contact constraint violation
$g_\mathrm{n} = \delta_\mathrm{c}^{e_\mathrm{H}}$, with exponent $e_\mathrm{H} = 5/4$. For the Hertz contact model, the constraint is assigned a compliance $\varepsilon_\mathrm{n} = e_\mathrm{H} / k_\mathrm{n}$ that is related to the stiffness coefficient $k_\mathrm{n} = E*\sqrt{d^*}/3$ with effective Young's elasticity modulus $E^* = [(1-\nu^2_a)/E_a + (1-\nu^2_b)/E_b]^{-1}$, Poisson ratio $\nu$, and effective diameter $d^* = (d_a^{-1} + d_b^{-1})^{-1}$ for the two contacting particles $a$ and $b$. Alternatively, a linear stiffness model is realized with $e_\mathrm{H} = 1$.
High-velocity impacts are modeled using the Newton impact law while preserving all other kinematic constraints. For resting contacts (low relative normal contact velocity), the damping coefficient is set $\gamma_\text{n} = 4.5 \dt$ to critically damp relative motion in the normal direction within a few timesteps $\dt$.
More details about the mapping and parametrization of the contact model and the precise Jacobians are found in \cite{servin:2014:esn,wiberg2021discrete}.

The dynamics system is time-integrated using the SPOOK stepper \cite{Lacoursiere2007}, which is a first-order accurate discrete variational integrator developed particularly for fixed timestep real-time simulation of multibody systems with non-ideal constraints and non-smooth dynamics. The time-discrete equations, forming a mixed complementarity problem (MCP), are solved using the direct-iterative split solver in AGX \cite{AGX2024}. 
A block-sparse LDLT solver with pivoting \cite{lacoursiere2010} is used as direct solver
for articulated rigid multibody systems and their contacts, with linearization of the Coulomb friction model. The MCP equations for DEM systems, and their coupling to multibody systems, are solved using a projected Gauss-Seidel (PGS) algorithm, which is accelerated using domain decomposition for parallel processing and warm-starting \cite{wang2016warm}. The PGS algorithm solves for the nonlinear Coulomb law and rolling friction laws, Eq.~(\ref{eq:G_t}) and (\ref{eq:G_r}), i.e., using no linearization. Setting an appropriate number of PGS solver iterations, $N_\text{it}$, are important. The convergence rate depends on the contact network and a guiding rule \cite{servin:2014:esn} is
\begin{equation}\label{eq:Nit}
    N_\text{it} = 0.1 n_d / \epsilon
\end{equation}
where $\epsilon$ is the error tolerance and $n_d$ is the length of the contact network (number of particles)
in the direction of the dominant stress. A consequence of the time-implicit nature of nonsmooth DEM is that the timestep is not restricted by the natural contact time period $\sqrt{m/k_\mathrm{n}}$. Instead, we use a similar guiding rule as in \cite{servin:2014:esn}
\begin{equation}\label{eq:timestep}
    \dt = \tfrac{1}{2}\sqrt{\epsilon d/a}
\end{equation}
where $a$ is the largest acceleration that can occur from the forces acting on a particle, which we estimate by $a \sim \sigma A / m = 3\sigma /2\rho d$ using the normal stress $\sigma$, particle cross-section area $A$, mass and $m$, and specific mass density $\rho$.

In the case of uniform particle beds of height $h$, the number of PGS iterations in Eq.~(\ref{eq:Nit}) is simply estimated using $n_d = h/d$ and the timestep from $\dt = d \sqrt{\epsilon\rho/6\sigma}$. When applying particle refinement in the vertical direction, this changes to
\begin{align}\label{eq:contact_network}
    n_d & = \eta \frac{\log (d_\text{max}/d_\text{min})}{\log r} + \tfrac{h - z_\text{max}}{d_\text{max}} 
\end{align}
where the first term is the number of particles in the vertical direction of the refined layers and the second term is the number of particles in the uniform bottom region starting at depth $z_\text{max}$.

\section{Simulations}
\label{sec:simulations}
In this section, we describe the simulation procedure for testing the effect of particle refinement on the mechanical response of digital soil samples and on the computational intensity. Triaxial tests and combined pressure-sinkage and shear-displacement tests are carried out with soil samples with different scaling aggressiveness and with different nearly uniform particle sizes for comparison. 

\subsection{Generation of soil samples}
\label{sec:bedGen}
The digital soil samples are meant to represent dry sand, and we follow \cite{servin2021multiscale} in using the following parameters for the spherical (nonsmooth) DEM particles: specific mass density $2200$ kg/m$^3$, Young's elasticity $1$ GPa, Poisson ratio $0.15$, friction coefficient $\mu_\text{t} = 0.3$, and rolling resistance coefficient $\mu_\text{r} = 0.02$.  
Cohesion is set to zero. The soil samples are created by emitting particles to fill up a rectangular container from above with normal earth gravity pointing downwards. The particles gradually form a bed and are left to relax until fully settled. During the emission and relaxation phase, both friction and rolling resistance are reduced to $\mu_\text{t} = 0.1$ and $\mu_\text{r} = 0.01$ respectively, with zero at the confining walls, to ensure that the sample is homogeneous and densely packed.

When creating granular beds with a desired particle refinement profile $d(z)$, we do this in a finite number of layers from the bottom and up. When each layer $n$ with particle diameter $d_n$ is filled to a height $\eta d_n$, the emitter is paused, and any excessive particles are removed before the emitter resumes with the next layer.
To avoid the formation of regular structures, the particle size within each layer is uniformly randomized in the range of $0.9 d_n$ to $1.1 d_n$. This size distribution applies to all layers, including $d_\text{min}$ and $d_\text{max}$, and also to uniform beds.

If particle refinement scaling is too aggressive, $\gamma \sim 1$, we observe notable size mixing between the layers. These cases are therefore excluded from the simulations. After generation, each bed was checked for internal mass density (or void ratio) variations. Within each bed, the mass density averaged $1400$ kg/m$^3$, with local variations of $1 - 7\%$ from this mean value, depending primarily on the size of the coarsest particles and less on the scaling aggressiveness.

\subsection{Triaxial test and invariance under particle refinement}
\label{sec:triaxial}
To verify that the bulk-mechanical properties of the DEM model is invariant under particle refinement, we conduct a series of triaxial tests simulations. These tests use a box-shaped consolidated drained triaxial test following the configuration and procedure described in \cite{wiberg2021discrete}. The box width is $w = 300$ mm for all tests and the particle-wall friction and rolling resistance are set to zero throughout the test.

We perform triaxial tests with seven different particle size configurations. For the soil samples with nonuniform particle size, we test with two different gradients (scaling aggressiveness), $\gamma = 0.050$ and $0.072$. In the tests, the particle diameter ranges from either $d_\text{min} = 8.5$ mm or $15$ mm to $d_\text{max} = 30$ mm, respectively, between opposing sidewalls.
Size gradients in both the vertical (axial) and horizontal directions are tested for. In the case of uniform particle sizes ($\gamma = 0$), triaxial tests are performed using three different particle sizes, namely $d = 8.5$ mm, $15$ mm, and $30$ mm, with a narrow size randomization as described in Sec.~\ref{sec:bedGen}. 

The tests are repeated with three different consolidation stresses on the horizontal sidewalls, $\sigma_2 = \sigma_3 = 5$ kPa, $15$ kPa, and $25$ kPa, while recording the vertical stress $\sigma_1$ as the system is forced to shear at a displacement rate of $5$ mm/s in the axial direction. The bulk internal friction and cohesion are computed use the Mohr-Coulomb failure criterion by fitting the Mohr circles with minor stress $\sigma_2 = \sigma_3$ and peak strength as the deviatoric stress $\sigma_{dev} = \sigma_1 - \sigma_3$.
Each test is repeated three times.

The resulting internal friction is presented in Table~\ref{tab:triaxial}. The results are invariant to particle size and size gradient within an error tolerance of $3\%$.

\begin{table}[htbp]
\centering
\caption{Results of triaxial tests with different particle sizes, uniform and particle refinement in vertical and horizontal direction.}
\label{tab:triaxial}
\begin{tabular}{lcl} 
\toprule
Test & Diameter(mm) & Friction Angle \\
\midrule
\multicolumn{3}{l}{\footnotesize{Uniform size ($\gamma = 0$)}} \\ \cmidrule(lr){1-3}
1 & 8.5 & 34.3 ($\pm$ 0.3)° \\
2 & 15 & 34.9 ($\pm$ 0.4)° \\
3 & 30 & 33.2 ($\pm$ 1.0)° \\
\midrule
\multicolumn{3}{l}{\footnotesize{Vertical gradient $\gamma = 0.050, 0.072$}} \\ 
\cmidrule(lr){1-3}
4 & $8.5\rightarrow 30$ & 34.7 ($\pm$ 0.4)° \\
5 & $15\rightarrow 30$ & 33.9 ($\pm$ 0.3)° \\
\midrule
\multicolumn{3}{l}{\footnotesize{Horizontal gradient $\gamma = 0.050, 0.072$}}\\ 
\cmidrule(lr){1-3}
6 & $8.5\rightarrow 30$ & 33.7 ($\pm$ 0.9)° \\
7 & $15\rightarrow 30$ & 33.5 ($\pm$ 1.0)° \\
\bottomrule
\end{tabular}
\end{table}

\subsection{Pressure-sinkage and shear-displacement}
\begin{figure}[h]
    \centering
    \includegraphics[width=1.0\linewidth]{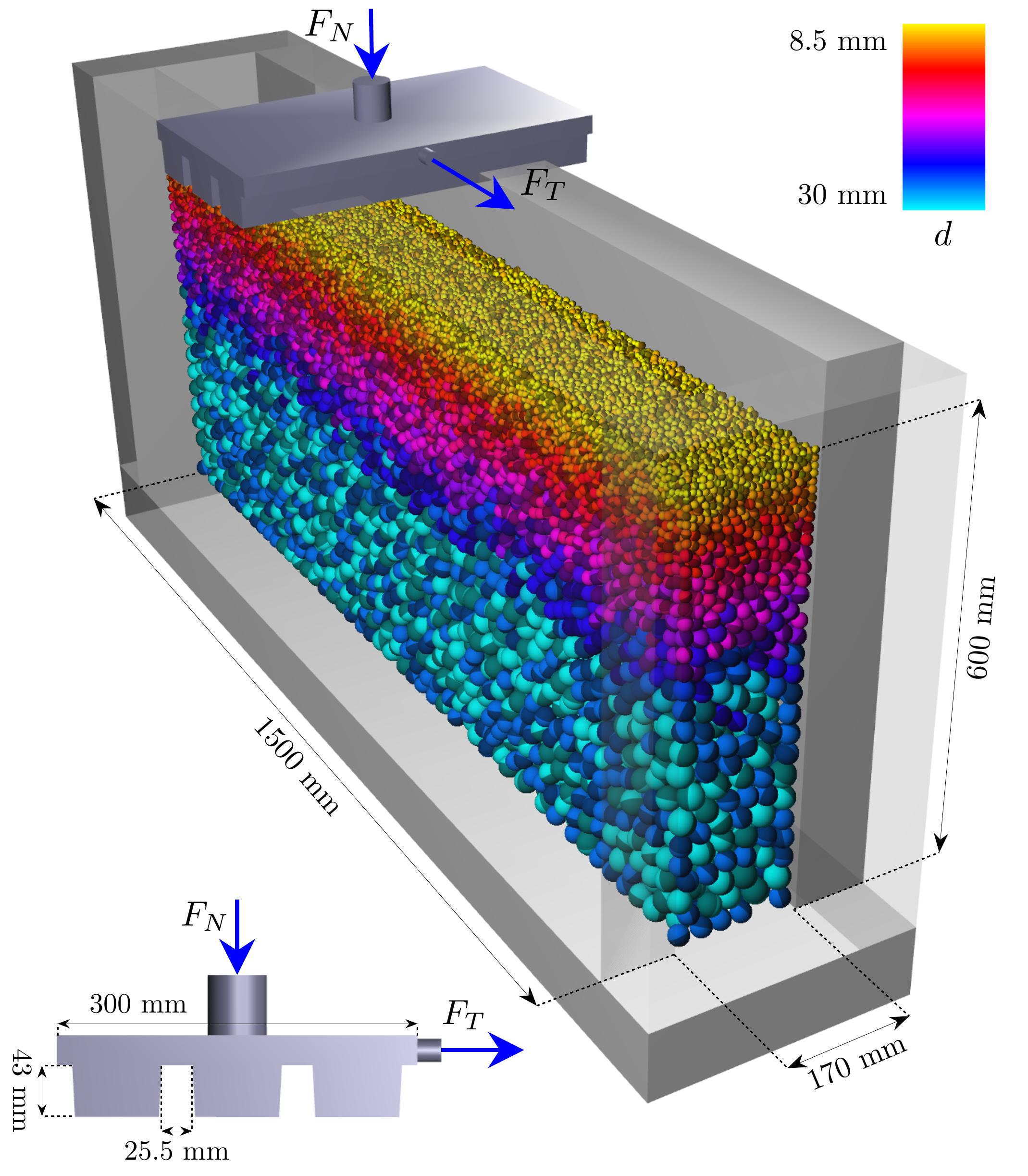}
    \caption{The container and the plate with their dimensions. Particles are coloured by their diameter.}
    \label{fig:container-track}
\end{figure}

To evaluate the effectiveness and accuracy of particle refinement in a more typical terramechanics application, we simulate a combined pressure-sinkage and shear-displacement tests. These tests are chosen as they effectively measure the performance of tractive elements depending on their shape and soil strength. For practical reasons, the two tests are carried out in a continuous sequence, i.e., the shear-displacement test is applied on the system state that is the result of the pressure-sinkage test. The setup is displayed in Fig.~\ref{fig:container-track}

The plate is 300 mm long with grousers that are 43 mm deep and 25.5 mm long at the bottom. The soil bed is narrower than the plate is wide ($500$ mm), i.e., the bed represents a thin slice of a larger domain to be simulated in areal application. The bed is 600 mm high, 1500 mm long, and 170 mm wide. We run simulations with 36 different configurations for particle refinement. 
In all cases, they have particles of size $d_\text{min} = 8.5$ mm at the top and then different scaling aggressiveness down to a depth, $z_\text{max}$, where $d_\text{max}$ is either 15, 22.5 or 30 mm. Below that depth, the particles have uniform size of $d_\text{max}$. The precise bed configurations and simulation settings are shown in Table~\ref{tab:bed-configurations-refinement}.
For comparison, we run simulations with uniform particle sizes set to 8.5, 15, 22.5 and 30 mm. See Table~\ref{tab:bed-configurations-uniform} for the configurations.  Each case is repeated five times, including the bed generation phase. This amounts to a total of 200 simulations. 

\begin{table*}[t]
\centering
\caption{The 36 bed configurations for particle refinement simulations. The lower value of scaling aggressiveness, $\gamma$, matches the higher values for the number of layers, particles, and solver iterations.}
\label{tab:bed-configurations-refinement}
\begin{tabular}{ccccccc}
\toprule
$d_\text{min}$ [mm] & $d_\text{max}$ [mm] & $\gamma$ & Num. layers & Num. particles & $N_\text{it}$ & $dt$ [ms] \\
\midrule
8.5 & 15 & 0.012 -- 0.37 & 4 -- 27 & 61k -- 134k & 215 -- 275 & 0.10 \\
8.5 & 22.5 & 0.025 -- 0.70 & 4 -- 22 & 22k -- 80k & 215 -- 275 & 0.10 \\
8.5 & 30 & 0.037 -- 0.96 & 4 -- 19 & 12k -- 57k & 215 -- 275 & 0.10 \\
\bottomrule
\end{tabular}
\end{table*}

\begin{table}[h]
\centering
\caption{Bed configurations for uniform particle size simulations.}
\label{tab:bed-configurations-uniform}
\begin{tabular}{ccccc}
\toprule
$d$ [mm] & $\gamma$ & Num. particles & $N_\text{it}$ & $dt$ [ms] \\
\midrule
8.5 & 0 & 305k & 375 & 0.10 \\
15.0 & 0 & 56k & 225 & 0.18 \\
22.5 & 0 & 17k & 175 & 0.25 \\
30.0 & 0 & 7k & 125 & 0.35 \\
\bottomrule
\end{tabular}
\end{table}


The sidewalls are frictionless. The plate-particle friction and rolling resistance coefficients are identical to the interparticle ones, listed in Sec.~\ref{sec:bedGen}. 

In the pressure-sinkage test, a vertical load of $F_N = 2550$ N is applied to a plate with grousers interfacing a horizontal soil bed while measuring the static sinkage and normal force on the plate over 2.5 s. We refer to this as phase I. The load corresponds to a normal stress of 50 kPa over the whole plate area. During pressure-sinkage, only the grousers are touching the soil, and the resulting sinkage is around 1-4 mm, depending on the bed. In the shear-displacement test, the plate is pulled horizontally while maintaining a constant vertical load force $F_N$ and registering the required pull force, $F_T$, as well as the dynamic sinkage. This allows us to determine the plate's tractive resistance, $F_T/F_N$, and how this evolves with the sinkage during shearing. The shear-displacement test starts with an acceleration phase, where the target pull speed is increased linearly from 0 to 100 mm/s over 0.1 s. At this speed, the inertial number $I$ is approximately $10^{-2}$, indicating dense flow conditions \cite{da2005rheophysics}. The bed resists shearing until it fails at some peak pull force. We refer to this as phase II and associate the time window 2.5 s and 3 s with it. As the plate is further displaced horizontally with the speed of 100 mm/s, the particle bed shears and the plate sinkage grows. We refer this as phase III and it lasts between time 3 s to the end of the test at 5.4 s. 

\begin{figure}[h]
    \centering
    \includegraphics[width=1.0\linewidth]{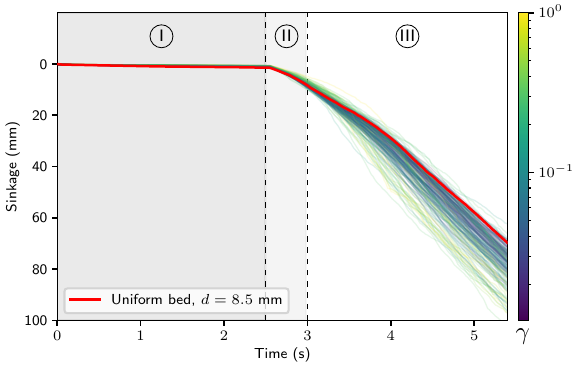}
    \caption{Track sinkage in simulations using particle beds with different scaling aggressiveness. The red curve is the reference case for uniform bed $d = 8.5$ mm.}
    \label{fig:sinkage}
\end{figure}

\begin{figure}[h]
    \centering
    \includegraphics[width=1\linewidth]{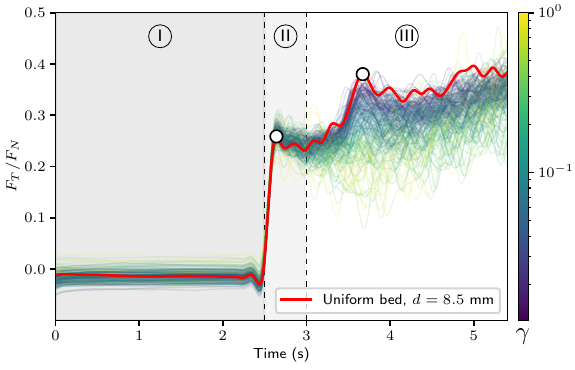}
    \caption{Coefficient of resistance, $F_T/F_N$, with the identified region marked. In the three later zones, a chosen peak is marked in each region. The red curve is the reference case for uniform bed $d = 8.5$ mm.}
    \label{fig:traction}
\end{figure}

The sinkage and traction curves from all simulations are shown in Fig.~\ref{fig:sinkage} and \ref{fig:traction}, respectively. The red line represents the reference case using a uniform particle size of $d = 8.5$ mm, while the background curves show results from different scaling aggressiveness ($\gamma$), with lower values of $\gamma$ producing results closer to the reference case. The general trend is that phase I and II are not very sensitive to scaling aggressiveness, but phase III shows deeper sinkage and lower traction with higher scaling aggressiveness. In the next section, we present a systematic analysis of this.

\begin{figure*}[t]
    \centering
    \includegraphics[width=\textwidth]{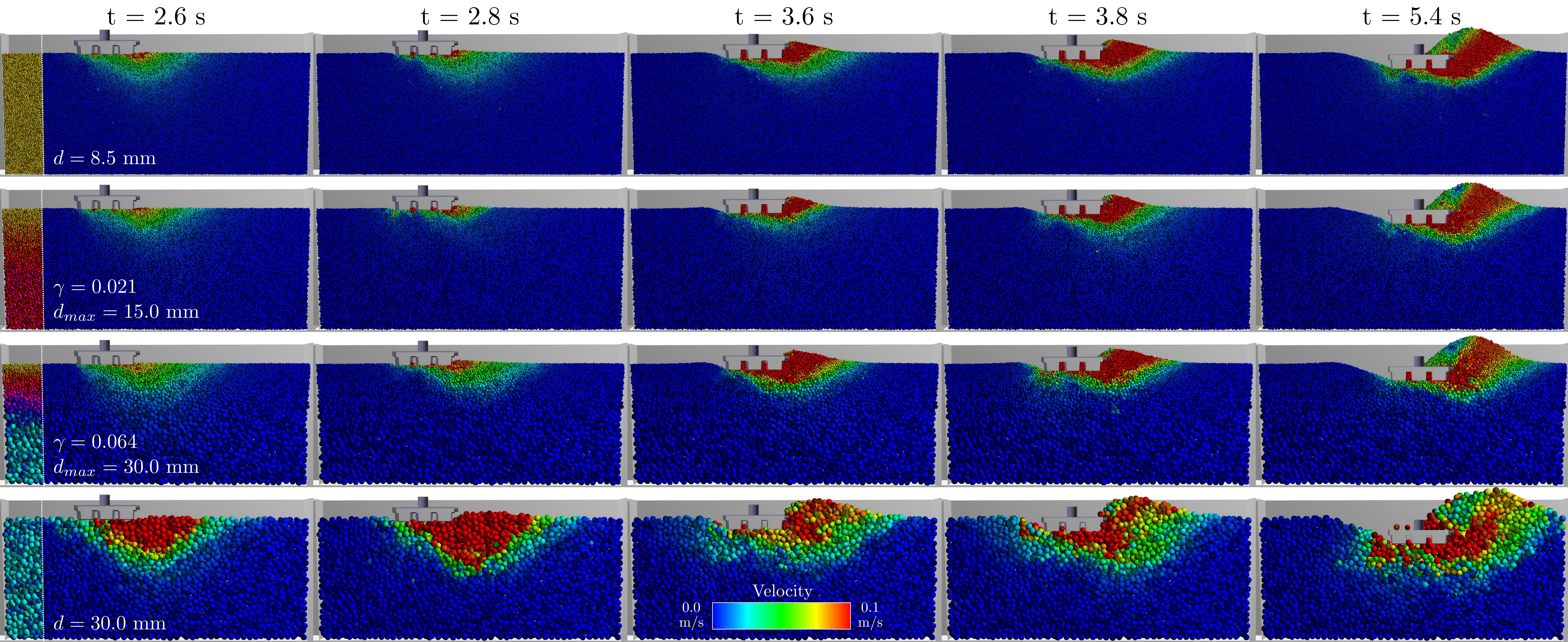}
    \caption{Sample images of particle velocities at fixed times during Phase II and Phase III. Particle colors indicate velocity magnitude according to the color bar (center bottom). The leftmost slice of the figure shows particle size distribution using a separate color gradient. From top to bottom: uniform bed with $d = 8.5$ mm, refined beds with $\gamma = 0.021$ and $\gamma = 0.064$, and uniform bed with $d = 30$ mm.}
    \label{fig:particle-velocities}
\end{figure*}

Sample images of particle velocities during phase II and III are shown in Fig.~\ref{fig:particle-velocities} for four soil beds with different particle sizes and scaling aggressiveness. The figure combines visualization of particle velocity magnitudes with a leftmost slice indicating particle sizes for each bed configuration. It is particularly evident that the particle flow and plate sinkage in the bed with the largest particles ($30$ mm) deviates considerably from the others. While these static images provide a snapshot of the behavior, the full dynamics can be better observed in the supplementary videos (see Supplementary Materials section).

\section{Results}
\label{sec:results}
The sinkage and traction behavior of 36 different particle refinement beds were analysed, with scaling aggressiveness, $\gamma$, ranging from $0.012$ to $0.96$ over three different $d_{max}$, $15$ mm, $22.5$ mm and $30$ mm. For analysis purposes, these beds were grouped into three ranges: low ($\gamma = 0.00$-$0.05$, $n=16$ beds), medium ($\gamma = 0.05$-$0.30$, $n=14$ beds), and high ($\gamma = 0.30$-$1.00$, $n=6$ beds). The particle refinement beds were compared to four different control beds of uniform particle size, with particle size $d$ as $8.5$ mm, $15$ mm, $22.5$ mm and $30$ mm. The control bed with $d = 8.5$ mm is our highest resolution uniform bed, and serves as our reference bed. 

Both the bed generation and the terramechanical simulations were repeated $5$ times for each configuration. These repetitions establish statistical significance for our analysis, with error bars in Fig.~\ref{fig:sinkage-I-III} and~\ref{fig:traction-peak},~\ref{fig:traction-avg} showing the variance between identical simulations run with different random seeds. This approach allows us to account for the inherent variability in granular systems while maintaining focus on the primary trends in the data.

\subsection{Sinkage}
\begin{figure}[h]
    \centering
    \includegraphics[width=1\linewidth,trim={6pt 6pt 6pt 6pt},clip]{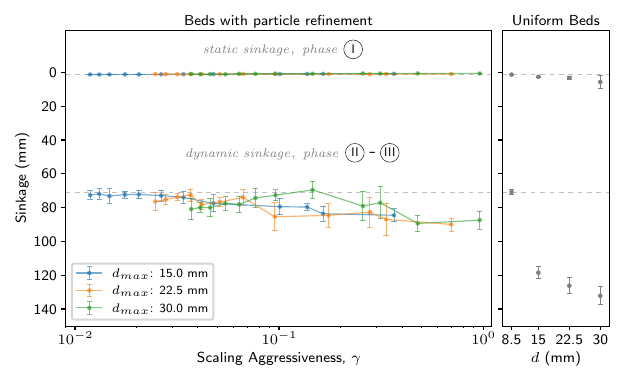}
    \caption{Sinkage in both phase I and phase II-III depending on particle size and scaling.}
    \label{fig:sinkage-I-III}
\end{figure}

Both the static sinkage during phase I and the dynamic sinkage during phase II-III were analysed for all beds. Fig. \ref{fig:sinkage-I-III} shows both of these sinkage measures for all scaling aggressiveness and also for the uniform control beds for comparison.

Static sinkage measurements showed a baseline value of $1.26$ mm for the high-resolution reference bed ($d = 8.5$ mm). Deviations from this reference varied systematically across the three $\gamma$ ranges. For low $\gamma$, the relative deviations were $17.4\%$ ($0.22$ mm absolute). In the medium range, the deviations increased to $33.8\%$ ($0.43$ mm absolute), while for high $\gamma$, deviations increased further to $44.1\%$ ($0.55$ mm absolute). In comparison, uniform beds demonstrated substantially larger deviations, ranging from $98.2\%$ for $d = 15.0$ mm to $349\%$ for $d = 30.0$ mm ($1.24$ mm to $4.38$ mm absolute). Notably, while these relative deviations appear significant, the absolute deviations remained small in all cases. Even the largest deviation of $4.38$ mm in uniform beds remained below the diameter of the smallest particles ($8.5$ mm) used in this study.

Dynamic sinkage measurements for the reference bed showed a baseline value of $70.8$ mm. For refinement beds, deviations increased systematically with $\gamma$, starting at $6.2\%$ ($4.4$ mm absolute) in the low range. These increased to $11.0\%$ ($7.8$ mm) for medium $\gamma$, and reached $21.3\%$ ($15.1$ mm) for high $\gamma$. Uniform beds exhibited much larger increases in deviation, ranging from $67.5\%$ ($47.8$ mm) at $d = 15.0$ mm to $86.7\%$ ($61.4$ mm) at $d = 30.0$ mm. While high $\gamma$ showed notable deviations from the reference, these remained significantly lower than those observed in any uniform bed configuration.

\subsection{Traction}
\begin{figure}[h]
    \centering
    \includegraphics[width=1\linewidth,trim={6pt 6pt 6pt 6pt},clip]{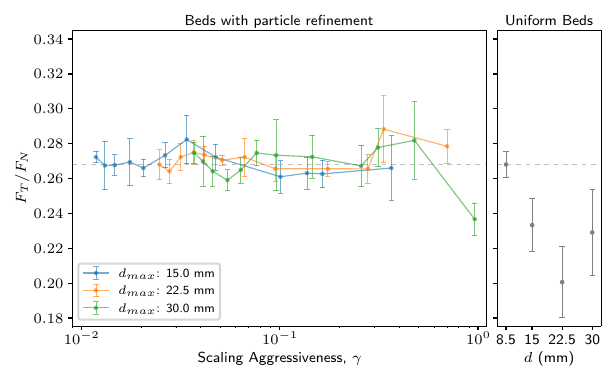}
    \caption{Initial peak traction in phase II depending on particle size and scaling.}
    \label{fig:traction-peak}
\end{figure}

\begin{figure}[h]
    \centering
    \includegraphics[width=1\linewidth,trim={6pt 6pt 6pt 6pt},clip]{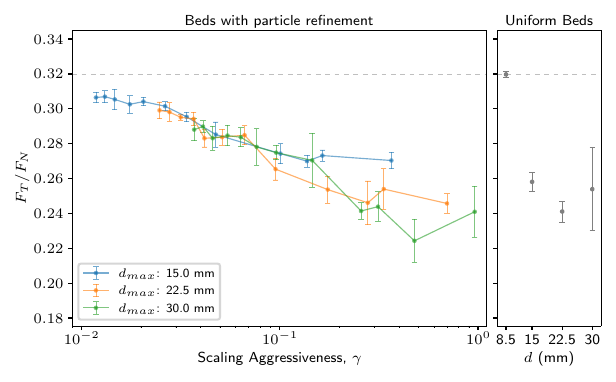}
    \caption{Average traction in phase II-III depending on particle size and scaling.}
    \label{fig:traction-avg}
\end{figure}

The traction behavior was analyzed in terms of both the initial peak traction during phase II (Fig.~\ref{fig:traction-peak}) and the average traction during phases II-III (Fig.~\ref{fig:traction-avg}). Both measures were evaluated across all refinement beds and compared to the uniform control beds.

The initial peak traction for the reference bed ($d = 8.5$ mm) showed a baseline value of $0.268$. For beds with low and medium $\gamma$, deviations remained remarkably small at $1.5\%$ and $1.6\%$, respectively, with values oscillating both above and below the reference. For high $\gamma$, an increase in deviation to $5.5\%$ was observed. Uniform beds, on the other hand, showed deviations between $13.0\%$ and $25.2\%$, with the largest deviation observed for $d = 22.5$ mm.

The average traction showed different characteristics compared to the initial peak values. The reference bed exhibited a baseline value of $0.320$. Refinement beds showed a systematic decrease in traction with increasing $\gamma$. Even at low $\gamma$, notable deviations of $7.4\%$ from the reference could be seen. These deviations increased to $15.5\%$ for medium $\gamma$, and reached $22.9\%$ for high $\gamma$. Uniform beds showed deviations ranging between $19.3\%$ and $24.6\%$, values similar to those observed in high $\gamma$ scaling beds, indicating comparable behavior during sustained shearing.

\subsection{Aggregate error}
To easily evaluate and compare the particle refinement method's performance, we combine the four previously discussed measures into a single aggregate error. For meaningful comparison across different scales, we first define normalized errors for both sinkage and traction measurements. The normalized sinkage error is defined as:
\begin{equation*}
    \varepsilon_{sinkage} = \frac{z - z_{ref}}{h_{grouser}}
\end{equation*}
where $h_{grouser} = 43.0$ mm is the grouser depth used as the normalization factor. This formulation applies to both static and dynamic sinkage measurements. For traction, the normalized error is:
\begin{equation*}
    \varepsilon_{traction} = \frac{F_T - F_{T,ref}}{F_{N,max}}
\end{equation*}
where $F_{N,max} = 2550$ N corresponds to the maximum normal load of $50$ kPa. This normalization is applied to both initial peak and average traction measurements. The aggregate error is then computed as the arithmetic mean of these four normalized errors.

Using the aggregate error, we analyzed the performance across all bed configurations. The results reinforce the trends observed in individual measurements. For refinement beds, the normalized error increased systematically with $\gamma$, starting at $3.4\%$ in the low range ($\gamma = 0.00$-$0.05$). This error increased to $6.2\%$ for medium $\gamma$ ($0.05$-$0.30$), and reached $11.3\%$ for high $\gamma$ ($0.30$-$1.00$). Uniform beds exhibited substantially larger errors between $30.9\%$ and $39.2\%$, with errors increasing with particle size. This comprehensive error measure demonstrates that the particle refinement method maintains significantly better accuracy compared to uniform particle beds across all scaling ranges.

\begin{figure}[h]
    \centering
    \includegraphics[width=1\linewidth,trim={6pt 6pt 6pt 6pt},clip]{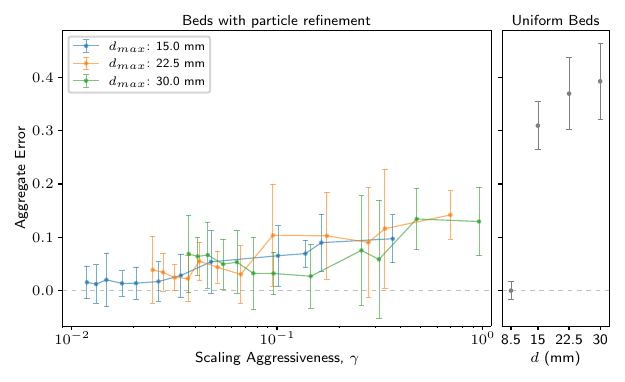}
    \caption{Aggregated error depending on particle size and scaling.}
    \label{fig:agg-error}
\end{figure}

\subsection{Number of particles and computational efficiency}
Finally, we analyze how the particle refinement affects the number of particles compared to the high-resolution reference case of approximately $\num{305000}$ particles. Even with the mildest application of particle refinement ($\gamma = 0.012$), the number of particles is reduced $2.3$ times ($56\%$ reduction), demonstrating immediate computational benefits. At the other extreme, the most aggressive scaling ($\gamma = 0.962$) reduces particle count $25$ times ($96.0\%$ reduction).

Across the three $\gamma$ ranges, particle reduction follows the expected trend. Low $\gamma$ configurations ($\gamma = 0.00$-$0.05$) average a $3.7$ times reduction, increasing to $7.9$ times for medium $\gamma$ ($0.05$-$0.30$), and reaching $13$ times for high $\gamma$ ($0.30$-$1.00$). For comparison, uniform beds show reductions of $5.5$ times at $d = 15.0$ mm, $19.0$ times at $d = 22.5$ mm, and $45.9$ times at $d = 30.0$ mm. Note also that uniform beds can achieve additional computational benefits through increased time steps as their smallest particle size increases (see Eq.~(\ref{eq:timestep}), where the error tolerance is kept constant).

While particle count reduction directly correlates with computational efficiency, the actual speed-up depends on the specific DEM implementation and solver strategy used. We observe better than linear speed-up with particle count reduction, as the particle refinement method also reduces the length of the contact network, allowing for fewer iterations in the Gauss-Seidel solver (see Eq.~(\ref{eq:Nit}) and~(\ref{eq:contact_network})). This results in simulation speed-ups ranging from $3.1$ to $43.0$ times faster compared to the reference case. However, these specific performance benefits are implementation-dependent and may vary across different DEM frameworks.

\begin{figure}[h]
    \centering
    \includegraphics[width=1\linewidth,trim={6pt 6pt 6pt 6pt},clip]{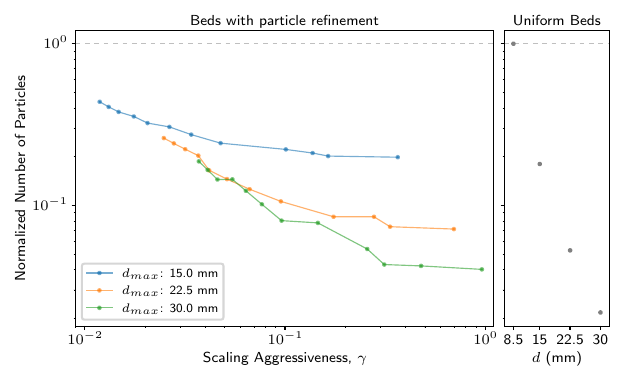}
    \caption{Effect of particle size and scaling aggressiveness on total particle count.}
    \label{fig:comp-intensity}
\end{figure}

\section{Discussion}
\vspace{1em}
\noindent\textit{Method limitations and applicability}
\vspace{0.5em}

The particle refinement method presented in this work was devised primarily for vehicle-terrain interaction scenarios. While our validation tests include cases with significant sinkage, demonstrating robustness to moderate soil deformation, certain aspects remain unexplored. For instance, as the plate width exceeds the container width, effects such as rutting could not be investigated. Also, for scenarios involving extensive soil displacement, such as excavation operations, the current refinement approach, being linear with depth, would require adaptation. For excavation, this could mean using radial refinement from the digging location, or maintaining the layer structure but with an extended top layer of fine particles. Another application can be found in ~\cite{mcdowell2012particle}, who employed radial refinement for cone penetrometer testing. Also, careful consideration must be given to particle size transitions, as smaller particles can percolate into the void spaces of larger particle zones. From our observations, low scaling shows no percolation, medium scaling shows occasional cases, and large scaling makes it more frequent.

These examples suggest that different applications demand different refinement strategies, requiring careful consideration of the specific scenario's deformation patterns.

These considerations suggest several directions for future development. In two dimensions, the current horizontal layer approach could evolve into half-circular arrangements around the area where the load is applied. Extended to three dimensions, this concept might employ hemispheric scaling zones. An even more ambitious development would be adaptive refinement, where a coarse particle background grid could be dynamically refined in regions of active interaction. Such a system could maintain computational efficiency by using fine particles only where needed, replacing them with larger particles in regions where detailed interaction is no longer required. A key challenge in implementing such dynamic refinement would be maintaining the integrity of force networks and stress distributions during particle size transitions.

\vspace{1em}
\noindent\textit{Comparison with bed depth truncation}
\vspace{0.5em}

A simpler alternative to particle refinement would be to simply truncate the simulation bed height, using only the upper portion of the domain with fine particles. Early in this work, we explored such domain scaling approaches, testing various reductions in length, width, and height. While height truncation might seem attractive due to its simplicity, experiments with bed heights ranging from 300 mm to 600 mm revealed significant limitations. Compared to full-height beds, we observed stress concentrations near the truncated bottom boundary, affecting the force network throughout the bed. The narrowly constrained domain was also associated with increased particle movement near the bottom, suggesting undesired boundary effects. While a 600 mm bed height might not be strictly necessary, our experiments indicated that maintaining at least 500 mm was important for accurate force propagation. Moreover, with the particle refinement method, the computational cost of the deeper regions remains low due to the larger particles while maintaining the necessary domain size for proper stress distribution, making height truncation unnecessary.

To illustrate the computational aspects: bed truncation from $600$ mm to $300$ mm would halve both the PGS solver iteration count, $N_\text{it}$, from $\sim\!400$ to $\sim\!200$, and reduce particle numbers from $\sim\!\num{300000}$ to $\sim\!\num{150000}$. While particle refinement, with $\gamma = 0.021$, achieves slightly better computational savings, reducing the particle count to $\sim\!100000$ and the iteration count to $\sim\!250$, while maintaining the full bed height necessary for proper stress distribution. 


\vspace{1em}
\noindent\textit{Importance of resolving grouser-particle interactions}
\vspace{0.5em}

During our investigations of grouser-particle interactions, we observed that having sufficiently small particles at the top surface played an important role in the overall behavior. Our observations indicate that when particles are small enough to fill the grouser cavities, there is an interlocking mechanism with the underlying particle layers. This interaction can enable the activation of a larger soil mass directly under the plate, where the surface particles can effectively engage with progressively larger particles in deeper layers.

This behavior can be seen in Fig.~\ref{fig:traction}, where the traction response shows a characteristic double-peak pattern in beds with small surface particles. After the initial peak and subsequent dip in traction force - occurring as the soil begins to shear - a second peak emerges. We suggest this secondary peak reflects the additional force required to mobilize an extended mass of soil, facilitated by the gradual particle size transitions from particle refinement. The small particles near the grouser initiate force chains that propagate through the particle size gradient, engaging a larger volume of soil than would occur with abrupt size transitions between layers. While this mechanism is difficult to visualize in the static snapshots of Fig.~\ref{fig:particle-velocities}, the dynamic behavior can be better observed in the supplementary videos (see \hyperref[sec:suppl]{Supplementary Material}).

\section{Conclusion}
The computational intensity of DEM simulations remains a significant challenge in terramechanics. This work demonstrates that a strategic particle refinement approach can effectively balance computational efficiency and accuracy by preserving fine particles in critical interaction zones while allowing larger particles elsewhere.

Compared to uniform high-resolution simulations, particle counts are reduced by factors ranging from 2.3 to 25, while normalized errors remain well-controlled between 3.4\% and 11\%. These results demonstrate that substantial computational savings can be achieved without significantly compromising simulation accuracy.

This approach contributes to achieving large-scale DEM simulations in terramechanics, enabling more extensive and complex analyses while maintaining acceptable computational costs. The method's effectiveness in preserving critical surface interactions while reducing overall particle count suggests its potential utility across various terramechanical applications.

\vspace{1.5em}
\noindent\textbf{CRediT authorship contribution statement}
\vspace{0.5em}

\noindent\textbf{Markus Pogulis}: Conceptualization, Methodology, Software, Formal analysis, Investigation, Data Curation, Writing – original draft, Writing – review \& editing, Visualization, Project administration. \textbf{Martin Servin}: Conceptualization, Methodology, Resources, Writing - Original Draft, Writing - Review \& Editing, Visualization, Supervision

\vspace{1.0em}
\noindent\textbf{Supplementary material}\label{sec:suppl}
\vspace{0.5em}

Simulation scripts, data, and supplementary figures and videos are available at: \url{https://umit.cs.umu.se/refinement/} and \url{https://doi.org/10.6084/m9.figshare.28159556.v1}


\vspace{1em}
\noindent\textbf{Declaration of competing interest}
\vspace{0.5em}

The authors declare that they have no known competing financial interests or personal relationships that could have appeared to influence the work reported in this paper.


\vspace{1em}
\noindent\textbf{Declaration of generative AI and AI-assisted technologies in the writing process}
\vspace{0.5em}

During the preparation of this work the author(s) used Anthropic's Claude 3.5 Sonnet in order to improve overall writing. After using this tool/service, the author(s) reviewed and edited the content as needed and take(s) full responsibility for the content of the publication.


\vspace{1em}
\noindent\textbf{Acknowledgements}
\vspace{0.5em}

The present work has in part been funded by BAE Systems Hägglunds AB, Mistra Digital Forest Grant DIA 2017/14 \#6, and Algoryx Simulation AB, which is gratefully appreciated. The computations were enabled by resources provided by the National Academic Infrastructure for Supercomputing in Sweden (NAISS), partially funded by the Swedish Research Council through grant agreement no. 2022-06725.



\end{document}